\documentclass[]{aa}
\usepackage{natbib,twoopt}
\usepackage[breaklinks=true,hidelinks,draft]{hyperref}
\bibpunct{(}{)}{;}{a}{}{,}
\usepackage{graphicx}

\makeatletter
  \newcommandtwoopt{\citeads}[3][][]{\href{http://adsabs.harvard.edu/abs/#3}%
    {\def\hyper@linkstart##1##2{}%
     \let\hyper@linkend\@empty\citealp[#1][#2]{#3}}}
  \newcommandtwoopt{\citepads}[3][][]{\href{http://adsabs.harvard.edu/abs/#3}%
    {\def\hyper@linkstart##1##2{}%
     \let\hyper@linkend\@empty\citep[#1][#2]{#3}}}
  \newcommandtwoopt{\citetads}[3][][]{\href{http://adsabs.harvard.edu/abs/#3}%
    {\def\hyper@linkstart##1##2{}%
     \let\hyper@linkend\@empty\citet[#1][#2]{#3}}}
  \newcommandtwoopt{\citeyearads}[3][][]%
    {\href{http://adsabs.harvard.edu/abs/#3}
    {\def\hyper@linkstart##1##2{}%
     \let\hyper@linkend\@empty\citeyear[#1][#2]{#3}}}
	\newcommandtwoopt{\citealtads}[3][][]{\href{http://adsabs.harvard.edu/abs/#3}%
    {\def\hyper@linkstart##1##2{}%
     \let\hyper@linkend\@empty\citealt[#1][#2]{#3}}}
\makeatother

\begin{document}
\title{Complex diffuse emission in the $z=0.52$ cluster PLCK~G004.5-19.5}
\author{J.\ G.~Albert\inst{1} \and C.~Sif\'on\inst{1,5} \and A.~Stroe\thanks{ESO Fellow}\inst{2,1}  \and F.~Mernier\inst{3,1} \and H.\ T.~Intema\inst{1} \and H.\ J.\ A.~R\"ottgering\inst{1} \and G.~Brunetti\inst{4}}

\institute{Leiden Observatory, Leiden University, P.O. Box 9513, NL-2300 RA Leiden, The Netherlands \and European Southern Observatory, Karl-Schwarzschild-Str. 2, 85748, Garching, Germany \and SRON Netherlands Institute for Space Research, Sorbonnelaan 2, 3584 CA Utrecht, The Netherlands\and INAF – Istituto di Radioastronomia, via P. Gobetti 101, 40129 Bologna, Italy\and Department of Astrophysical Sciences, Peyton Hall, Princeton University, Princeton, NJ 08544, USA}

\date{\today}
\abstract{
	We present radio observations of the galaxy cluster PLCK~G004.5-19.5 ($z=0.52$) using the Giant Metrewave Radio Telescope at 150~MHz, 325~MHz, and 610~MHz.
	We find an unusual arrangement of diffuse radio emission in the center and periphery of the cluster, as well as several radio galaxies with head-tail emission.
	A patch of peripheral emission resembles a radio relic, and central emission resembles a radio halo.
	Reanalysis of archival \emph{XMM-Newton} X-ray data shows that PLCK~G004.5-19.5 is disturbed, which has a known correlation with the existence of radio relics and halos. 
	Given that the number of known radio halos and radio relics at {$z>0.5$} is very limited, PLCK~G004.5-19.5 is an important addition to understanding merger-related particle acceleration at higher redshifts.
}
\maketitle
\section{Introduction}
Galaxy clusters are the most massive gravitationally bound structures in the Universe with masses and volumes of the order of $10^{14-15}~\mathrm{M}_\sun$ and $100~\text{Mpc}^{3}$.
Hierarchical mergers between clusters at the intersections of the cosmic web, with relative velocities near $1000~\mathrm{km}\,\mathrm{s}^{-1}$, can release gravitational energy of the order of $10^{63}~\mathrm{erg}$.
A fraction of this energy is dissipated into shocks and turbulence, which in turn accelerate cosmic ray electrons (CRe) and hadrons in the intracluster medium (ICM).
The shock acceleration might occur via diffusive shock acceleration \citepads[DSA; for a comprehensive review see][]{1983RPPh...46..973D,2001RPPh...64..429M,2013ApJ...764...95K}. 
Turbulence can reaccelerate relativistic particles via second-order Fermi mechanisms \citepads[for an updated review see]{2014IJMPD..2330007B}. 
The relativistic CRe give rise to synchrotron radiation in the presence of magnetic fields, which are typically of the order of $\mu\mathrm{G}$ \citepads{1999ApJ...518..594R}.

Emission tracing the shock fronts is referred to as radio relics, which are typically diffuse, elongated, and polarised sources located near the cluster periphery.
Central unpolarized, diffuse radio sources are known as radio halos and likely result from turbulent acceleration.
Clusters hosting radio relics and halos have a disturbed morphology \citepads{2010ApJ...721L..82C}, but not all merging clusters exhibit halos, suggesting that halos originate from a hierarchy of complex mechanisms \citepads{2013MNRAS.429.3564D,2016PPCF...58a4011B}.

The current catalogues of radio relics and halos are far from complete, mainly because of the limited sensitivities of existing radio surveys.
\citetads{2012MNRAS.420.2006N} 
performed an analysis of MareNostrum, a high-resolution cosmological simulation, and modelled the abundance of radio relics across redshift.
By assuming a DSA efficiency, they estimate $\gtrsim100$(800) to be found at $0.5<z<1$ by the upcoming Tier-1 survey of LOFAR \citepads{2013A&A...556A...2V} at 60~MHz (120~MHz).
Currently, there are only four relics observed beyond $z=0.5$ (\citeads[MACS~J1149.5+2223 at $z=0.54$][]{2012MNRAS.426...40B}; \citeads[MACS~J0717.5+3745 at $z=0.55$][]{2009A&A...503..707B,2009A&A...505..991V}; \citeads[El~Gordo at $z=0.87$][]{2014ApJ...786...49L,2016MNRAS.tmp.1213B} and \citeads[MACS~J0025.4-1222 at $z=0.586$][]{2017A&A...597A..96R}). 

The number of observed halos beyond $z=0.5$ has been limited by the sensitivity of low-frequency radio observations.
It is estimated that there are about 500--1000 radio halos for $z<0.6$ that should be found at 150~MHz by LOFAR Tier-1 \citet{2006MNRAS.369.1577C}.
This is because inverse Compton (IC) losses (scaling with $(1+z)^4$) become more efficient than DSA at higher redshifts \citepads{2006MNRAS.369.1577C}. 
Thus IC losses are able to suppress high-energy CRe, giving rise to an increasing fraction of radio halos with very steep spectra, which are only bright at low frequencies \citepads[e.g.][]{2008Natur.455..944B}.

One as yet unanswered question is how efficient DSA and turbulence are at electron acceleration.
In particular, it is unclear how the two mechanisms evolve over cosmic time, and what their impact is on the underlying magnetic field.
Furthermore, we eventually wish to understand how shocks and turbulence affect host galaxies following a merger.

We present galaxy cluster PLCK~G004.5-19.5, which lies at a redshift ($z=0.516$) with few known radio relics and halos; this makes it a tantalizing target to search for diffuse emission.
It was discovered by the \textit{Planck} satellite through the Sunyaev-Zel'dovich (SZ) effect and confirmed with X-ray observations \citepads{2011A&A...536A...9P,2014A&A...571A..29P}. 
This cluster is hot, $10.2\pm0.5$~keV, and very massive, $M^{SZ}_{500}=(10.4\pm0.7) \times 10^{14}~\mathrm{M}_\sun$, and it hosts strong lensing arcs \citepads{2014A&A...562A..43S}. 
Initial follow-up studies in low-resolution low-frequency archival data found that PLCK~G004.5-19.5 hosted strong radio emission \citepads{2014A&A...562A..43S}. 
However, since the radio sources were unresolved, the nature of the radio emission could not be solved. 

In this paper we offer new high-resolution 150~MHz, 325~MHz, and 610~MHz observations of the cluster PLCK~G004.5-19.5 using the Giant Metrewave Radio Telescope (GMRT), and a reanalysis of \textit{XMM-Newton} data for a morphology study of the X-ray emission. 
In Section~\ref{sec:obsandreduction} we explain our reduction method and how we corrected \textit{XMM-Newton} X-ray data of PLCK~G004.5-19.5 for background and vignetting.
In Section~\ref{sec:resultsboth} we analyse the X-ray and radio emission.

In this work, we assume a flat $\Lambda$CDM cosmology with $h=0.70$, and $\Omega_{\rm m}=0.30$, which gives an angular scale at $z=0.516$ of $6.2~$kpc/\arcsec.

\section{Observations and data reduction}
\label{sec:obsandreduction}
\subsection{GMRT observations of PLCK~G004.5-19.5}
\label{sec:obs}

We carried out a set of observations on PLCK~G004.5-19.5 using the GMRT at 150~MHz, 325~MHz (PI: A.\ Stroe, Project: 27\_051), and 610~MHz (PI: C.\ Sif\'on, Project: 25\_036) with total times on source of 445~min., 390~min., and 798~min.
We performed 15~min. observations on flux calibrators 3C~48 and 3C~286 at the start and end of the observation session, and 10~min. interleaved observations of the nearby and bright phase calibrator J1924-292 that has a flat spectrum amplitude \citepads{2007ApJS..171...61H}.


We removed the initial radio frequency interference (RFI) of the obvious corrupt time and frequency data by flagging by hand.
We then used the source peeling and atmospheric modelling pipeline \citepads[SPAM; for a full description see][]{2009A&A...501.1185I,2014ascl.soft08006I}.
SPAM performs antenna delay, bandpass, phase, and amplitude calibration, and then multiple rounds of self-calibration and further flagging of bad data. 
It then iteratively subtracts all sources except for bright calibrators, and calibrates direction dependently. 
It then spatially fits a model to the direction-dependent calibration products and derives interpolated ionospheric corrections.
 
For our choice of initial sky model for self-calibration we used a bootstrap method.
We first used the 150~MHz model from the TIFR GMRT Sky Survey Alternative Data Release (TGSSADR; \citetads{2016arXiv160304368I}) to start self-calibration for our 150~MHz observations, and then used our deeper 150~MHz model as the initial model for our 325~MHz self-calibration, likewise using the 325~MHz image for the 610~MHz initial sky model. 
The TGSSADR provides an excellent starting model for self-calibration in this lower frequency range and is available to the public\footnote{\url{http://tgssadr.strw.leidenuniv.nl/}}.

We performed multi-scale (\texttt{multiscale} = (0,4,8,32)~pixels) multi-frequency deconvolution with CASA on the SPAM-calibrated products with a slightly uniform Briggs weighting \citep[\texttt{robust}$=-0.3$,][]{briggs} and four pixels per beam.
We also corrected for a non-coplanar array using $w$-projection \citepads{2008ISTSP...2..647C}. 
The multi-scale clean avoids the issues related to representing resolved diffuse emission with a sum of point-like sources, and this choice of weighting balances beam uniformity in side lobes with beam resolution.
The highest minimum baseline of the three frequencies is $120~\lambda$, corresponding to a highest sensitivity scale of $\sim 20~\arcmin$. 

\begin{figure}[h]
\centering
	\resizebox{\columnwidth}{!}{\includegraphics{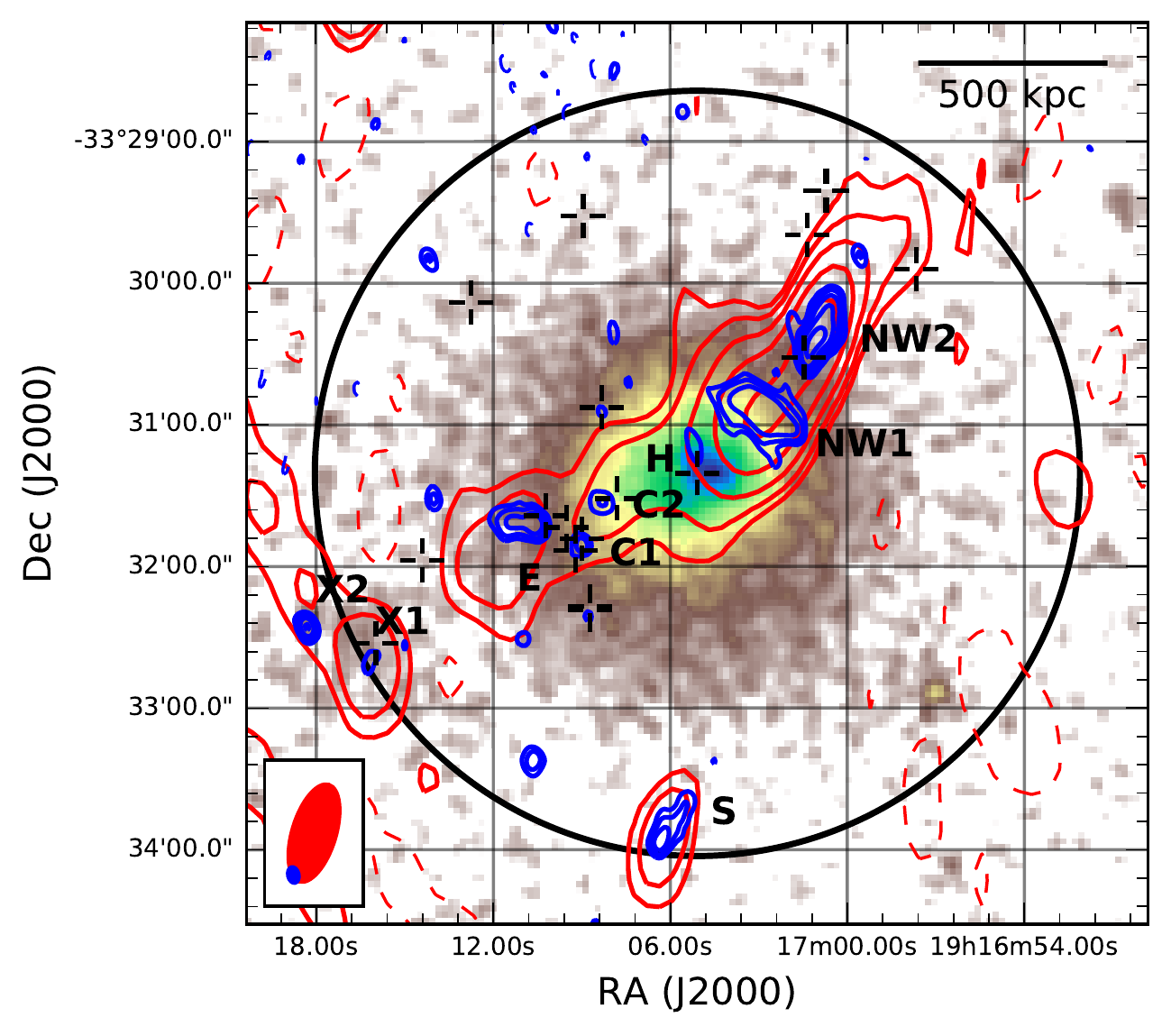}}
	\caption{150~MHz contours (red) and 610~MHz contours (blue) overlaid on the corrected \emph{XMM-Newton} EPIC image, with radio beams in the bottom left corner. 
	Contours are at $(5,10,20,40,80,160)\sigma_{\rm rms}$ levels.
    Thinner dashed contours mark $-5\sigma_{\rm rms}$ (they are only noticeable for 150~MHz).
	Black crosses show GMOS spectroscopic cluster members within the redshift range $z=0.516 \pm 0.015$. $R_{500}$ is shown by the black circle. 
    The 150~MHz beam is $43.3~\arcsec\times 18.9~\arcsec$, and 610~MHz is $7.2~\arcsec\times 4.9~\arcsec$.
	}
 \label{fig:radio150}
\end{figure}

\begin{figure}[h]
\centering
\resizebox{\columnwidth}{!}{\includegraphics{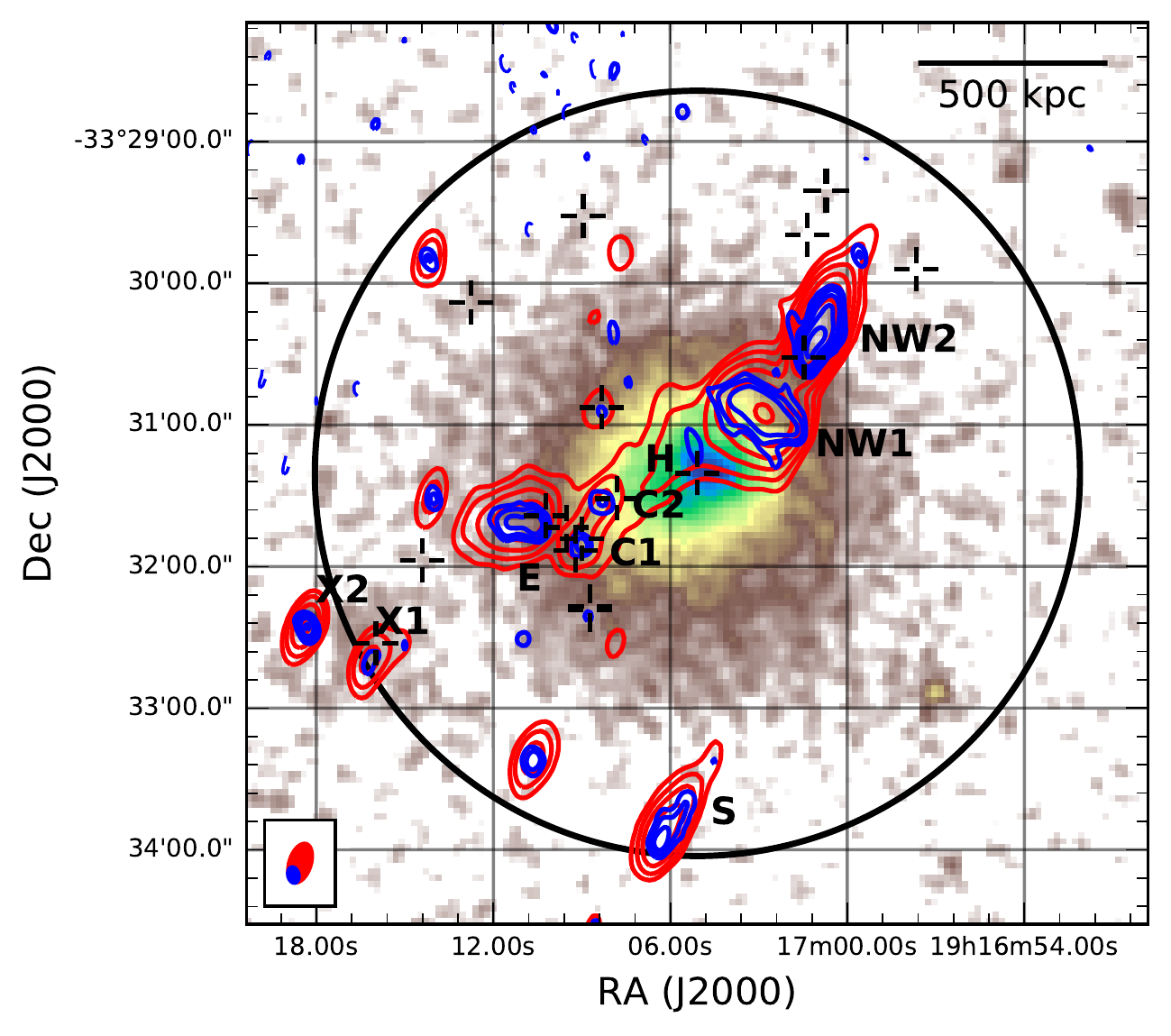}}
	\caption{Same legend as in Figure~\ref{fig:radio150} except that the red contours are 325~MHz. 
    The 325~MHz beam is $17.5~\arcsec\times 9.5~\arcsec$, and 610~MHz is $7.2~\arcsec\times 4.9~\arcsec$.}
 \label{fig:radio325}
\end{figure}

The background root-mean-squared noises for the individual radio maps (Figs.~\ref{fig:radio150} and \ref{fig:radio325}), $\sigma_{\rm rms}$, were calculated as the standard deviation of the residual after Gaussian source fitting and subtraction using Python Blob Detection and Source Measurement (PyBDSM) v1.8.
Our reduction yielded $\sigma_{\rm rms}=1.40$~mJy\,beam$^{-1}$ at 150~MHz, $\sigma_{\rm rms}=120$~$\mu$Jy\,beam$^{-1}$ at 325~MHz, and $\sigma_{\rm rms}=90$~$\mu$Jy\,beam$^{-1}$ at 610~MHz.
The resolutions of the maps are $43.3~\arcsec\times 18.9~\arcsec$, $17.5~\arcsec\times 9.5~\arcsec$, and $7.2~\arcsec\times 4.9~\arcsec$, respectively.
We note that the side lobes from a nearby bright source ($S_{1.4~{\rm GHz}}=230$~mJy from the NRAO VLA Sky Survey) cause a spatially varying background noise in our target field.

\subsection{\textit{XMM-Newton} reduction}
\label{sec:xmm}

PLCK~G004.5-19.5 was observed by \textit{XMM-Newton} on 23 March, 2010 (ObsID: 0656201001), for a total duration of 14.2~ks. 
We reduced the EPIC MOS\,1, MOS\,2, and pn data using the XMM Science Analysis Software (SAS) v14. 
We ran the standard pipeline tasks \texttt{emproc} and \texttt{epproc} for MOS and pn data, respectively. 
For each of the three detectors, we filtered the data for solar flare events, following the same method as described in \citetads{2015A&A...575A..37M}. 
In the 10--12~keV band of MOS and the 12--14~keV band of pn, we fitted count-rate histograms binned in 100~s intervals with a Poissonian curve. 
We excluded all the time intervals in which the count-rate lay above $\mu + 2\sqrt{\mu}$, where $\mu$ is the mean of the Poissonian distribution. 
We repeated the procedure for the 0.3--10~keV band in the three instruments with histograms binned in 10~s intervals, since \citetads{2004A&A...419..837D} reported that soft flare events can also affect soft X-ray bands. 
As recommended by the calibration reports, we kept only the highest quality events (\texttt{flag}$=0$) for the three instruments. 
Only single events were are allowed in pn (\texttt{pattern}$=0$), while we allowed single, double, triple, and quadruple events in MOS (\texttt{pattern}$\le12$).

We extracted the MOS\,1, MOS\,2, and pn images of PLCK~G004.5-19.5 within four distinct energy bands (0.3--2~keV, 2--4.5~keV, 4.5--7.5~keV, and 7.5--12~keV), and extracted similar ``background'' images, taken from stacked filter-wheel closed observations\footnote{\textit{XMM-Newton} SOC website (\texttt{http://xmm.esac.esa.int}).}. 
We scaled the average count-rate of the ``background'' images to the average count-rate of the EPIC images within 10--12~keV, where negligible cluster emission is expected. 
The EPIC images in each band, and for each instrument, were then subtracted from this instrumental background and corrected for vignetting effects, after dividing them by their respective exposure maps generated using \texttt{eexpmap}.

\begin{figure}[ht]
        \centering
			\resizebox{\columnwidth}{!}{\includegraphics[width=0.50\textwidth]{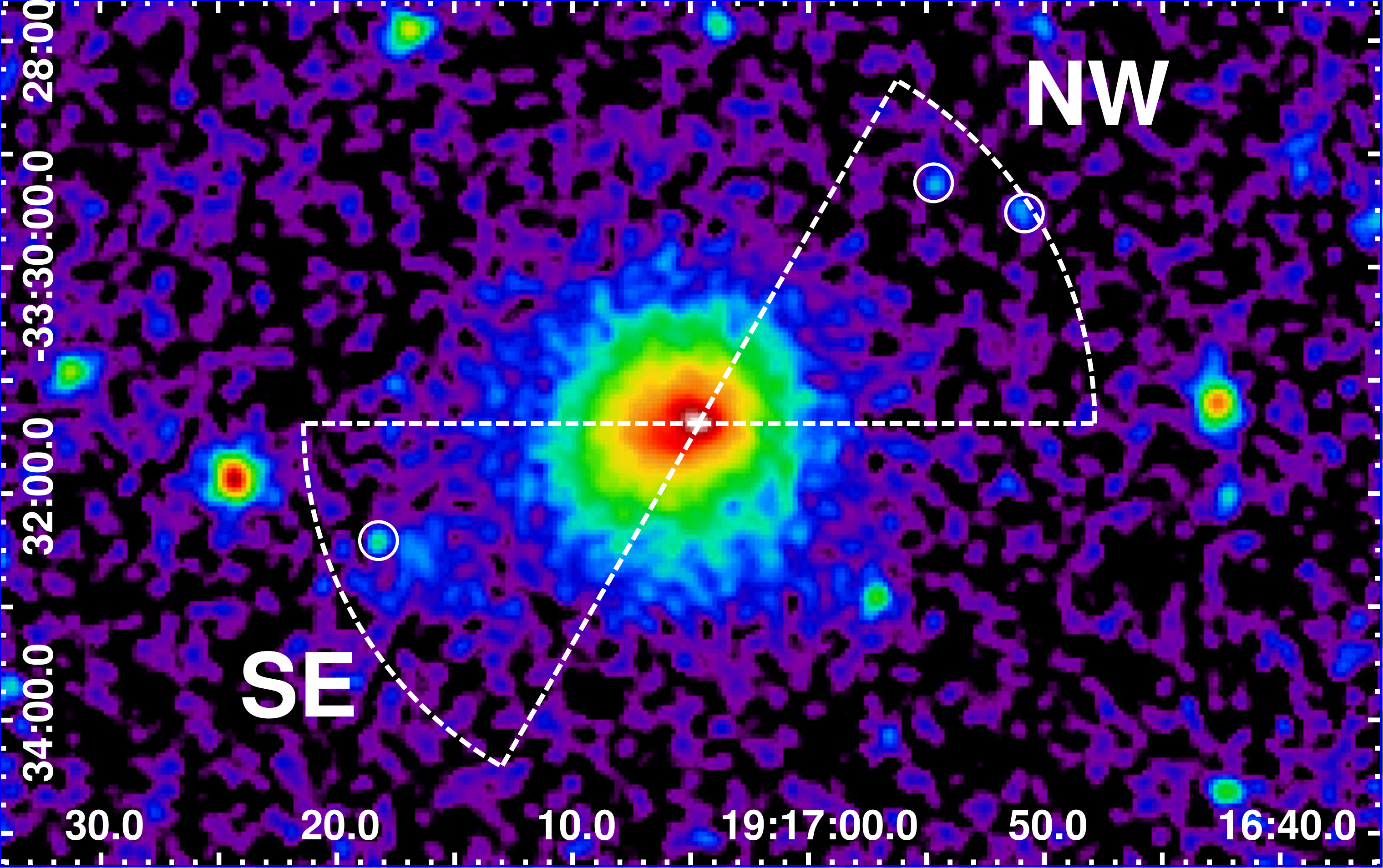}}
        \caption{\textit{XMM-Newton} EPIC (MOS+pn) background- and exposure-map corrected image of PLCK~G004.5-19.5 (0.3--12 keV). 
		The point-spread function of the EPIC cameras is $\approx 6~\arcsec$. 
		Overlaid are the extraction regions (wedges) and point-source masks (circles) used for the radial profile measurement.}
\label{fig:XMM_EPIC_image}
\end{figure}

The background- and vignetting-corrected images were then combined into one full EPIC image (Figs~\ref{fig:radio150}--Fig.~\ref{fig:XMM_EPIC_image}).

\subsection{Spectroscopic data}
We complement our X-ray and radio analysis with spectroscopic measurements of galaxy redshifts derived from two sets of observations. 
In addition to the six redshifts published by \citetads{2014A&A...562A..43S}, we retrieved spectroscopic observations performed with the Focal reducer and low-dispersion spectrograph (FORS2) on the Very Large Telescope (VLT; Program ID: 090.A-0925(A), PI: H.\ B\"ohringer) as part of a spectroscopic follow-up of \textit{Planck} $z>0.5$ cluster candidates. 
We reduced these data with the ESO \textsc{Reflex} pipeline, which performs standard reduction steps such as bias subtraction, flat-field correction and wavelength calibration \citepads{2013A&A...559A..96F}. 
We measured galaxy redshifts by cross-correlating the resulting spectra with template galaxy spectra from the Sloan Digital Sky Survey using the \textsc{rvsao} software \citepads{1998PASP..110..934K}. 
Combining there two data sets, we find 16 galaxies in the redshift range 0.50--0.53 (corresponding to a range of 3,000 $\mathrm{km}\,\mathrm{s}^{-1}$) that are isolated in redshift space and have a velocity dispersion of $\sim 1200~\mathrm{km}\,\mathrm{s}^{-1}$. The cluster redshift from the combined 16 galaxies is $z_\mathrm{cl}=0.519$.

\section{Results and discussion}
\label{sec:resultsboth}
\subsection{X-ray structure}
\label{sec:xmm_an}

Our processed EPIC image of PLCK~G004.5-19.5 (Fig.~\ref{fig:XMM_EPIC_image}) suggests that the cluster is disturbed. 
Assuming $z=0.516$, we find an offset $\sim100~\mathrm{kpc}$ between the X-ray peak 
and the X-ray centroid, 
estimated from a circular aperture of $R_{\rm ap}=500~\mathrm{kpc}$ on the X-ray peak.

The degree of cluster disturbance can be provided in a more quantitative way, by estimating at least two morphological parameters \citepads[described in detail in][]{2010ApJ...721L..82C}.

\begin{enumerate}

\item The centroid shift, $w$ \citepads{2006MNRAS.373..881P,2008ApJS..174..117M}, is estimated from fitting 2D $\beta$-profiles in a series of $N$ circular apertures $\xi_i R_\text{ap}$ (with $0.05 \le \xi_i \le 1$) centred on the X-ray peak, and can be expressed as
\begin{equation}
w = \frac{1}{R_\text{ap}} \times \bigg[  \frac{1}{N-1} \sum_{i=1}^{N} \, ( \Delta_i - \langle \Delta \rangle )^2 \bigg]^{1/2},
\end{equation}
where $\Delta_i - \langle \Delta \rangle$ is the difference between the X-ray peak and the centroid estimated in the $i$th aperture. 
Although \citetads{2010ApJ...721L..82C} proceeded from $0.05R_{\rm ap}$ to $R_{\rm ap}$ in 5\% steps, here the larger PSF of EPIC does not allow us to explore the image with such precision.
Therefore, here we chose larger aperture radii of $0.2R_{\rm ap}$, $0.4R_{\rm ap}$, $0.6R_{\rm ap}$, $0.8R_{\rm ap}$, and $R_{\rm ap}$.

\item The concentration parameter, $c$ \citepads{2008A&A...483...35S}, is defined as the ratio of the peak over the ambient surface brightness $S$:
\begin{equation}
	c = \frac{S(r<100~\mathrm{kpc})}{S(r<500~\mathrm{kpc})}.
\end{equation}

\end{enumerate}

Based on our EPIC image, we find $w = 0.058 \pm 0.007$ and $c = 0.079 \pm 0.004$. 
Referring to Fig. 1a of \citetads{2010ApJ...721L..82C}, it appears that PLCK~G004.5-19.5 is situated within the lower-right quadrant where disturbed galaxy clusters are found that host radio halos.
This suggests that PLCK~G004.5-19.5 is disturbed, although we note that PLCK~G004.5-19.5 lies further in redshift than all of the clusters used in the derivation of the relation.

We note a small nodule of previously unidentified X-ray emission to the south-east of the main X-ray profile in Fig.~\ref{fig:optxnod}.
A clustering of red elliptical galaxies, as well as a spectroscopic member, suggests that this nodule is at the cluster redshift.
Assuming the nodule is purely thermal emission, with $kT$ ranging from (3--10)~keV, we find unabsorbed luminosities in the (0.1--2.4)~keV energy band, within a circular region of $31\arcsec$ (192~kpc) and excluding a point-like source and main cluster X-ray emission\footnote{The south-east nodule falls between two CCD chips for the EPIC pn instrument, therefore we were unable to calculate its luminosity.},
\begin{align}
	L_X^{\rm MOS1} =& (6.7 \pm 1.4) \times 10^{43}~\mathrm{erg}\,s^{-1}\notag\\
	L_X^{\rm MOS2} =& (5.8 \pm 1.3) \times 10^{43}~\mathrm{erg}\,s^{-1}\notag,
\end{align}

These luminosities are consistent with typical values for small galaxy groups and sub-clusters \citepads{2002ApJ...567..716R}.
Taking the same aperture, $31\arcsec$, we calculate a main-cluster luminosity of $(9.2 \pm 1.4) \times 10^{44}~\mathrm{erg}\,s^{-1}$.
This is 14--16 times brighter than the luminosity of south-eastern nodule.

\begin{figure}[!]
        \centering
			\resizebox{\columnwidth}{!}{\includegraphics[width=0.50\textwidth]{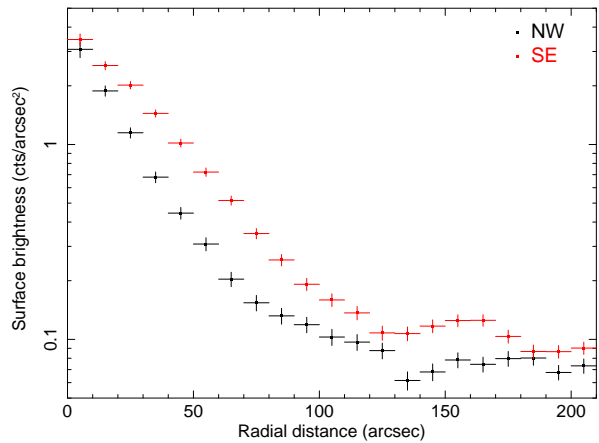}}
		\caption{Radial surface brightness profile of \textit{XMM-Newton} EPIC background- and exposure-map corrected image of PLCK~G004.5-19.5 (0.3--12)~keV in the NW and SE directions, centred on the X-ray brightness peak.}
\label{fig:SB_profile}
\end{figure}

In Fig.~\ref{fig:SB_profile} we show the radial X-ray (0.3--12)~keV surface brightness profile of PLCK~G004.5-19.5 in the south-east and north-west, centred on the X-ray brightness peak (see the mask in Fig.~\ref{fig:XMM_EPIC_image}).
We see a significant uniform over-brightness in the south-east relative to north-west.
This supports that the ICM is disturbed. 
We also note an excess located 160~\arcsec to the south-east that is associated with the X-ray nodule.

\subsection{Radio emission}
\label{sec:radio_an}

We are able to detect several diffuse and compact radio sources, which are labelled in bold in Figures~\ref{fig:radio150} and \ref{fig:radio325}.
We performed Gaussian source extraction using PyBDSM v1.8.7 with a detection threshold of $5\sigma_{\rm rms}$.
For point sources lacking a detection at one frequency, we measured the flux within an aperture of one beam.
For one resolved source, we measured the flux within a common aperture, and we discuss this procedure below.
The flux measurements are given in Table~\ref{tab:spectral}, with nondetection measurements labelled accordingly.

We take the flux variance to be the sum of the measurement variance and a systematic variance equal to 15\%\footnote{Suggestion of H.\ T.~Intema.} of the flux, that is, $\sigma_S^2 = \sigma_{\rm rms}^2A/A_{\rm beam} + (0.15 S)^2$, where $S$, $A$, and $A_{\rm beam}$ are the measured flux, source area, and beam area.

\begin{figure}[!]
        \centering
	\resizebox{1.0\columnwidth}{!}{\includegraphics{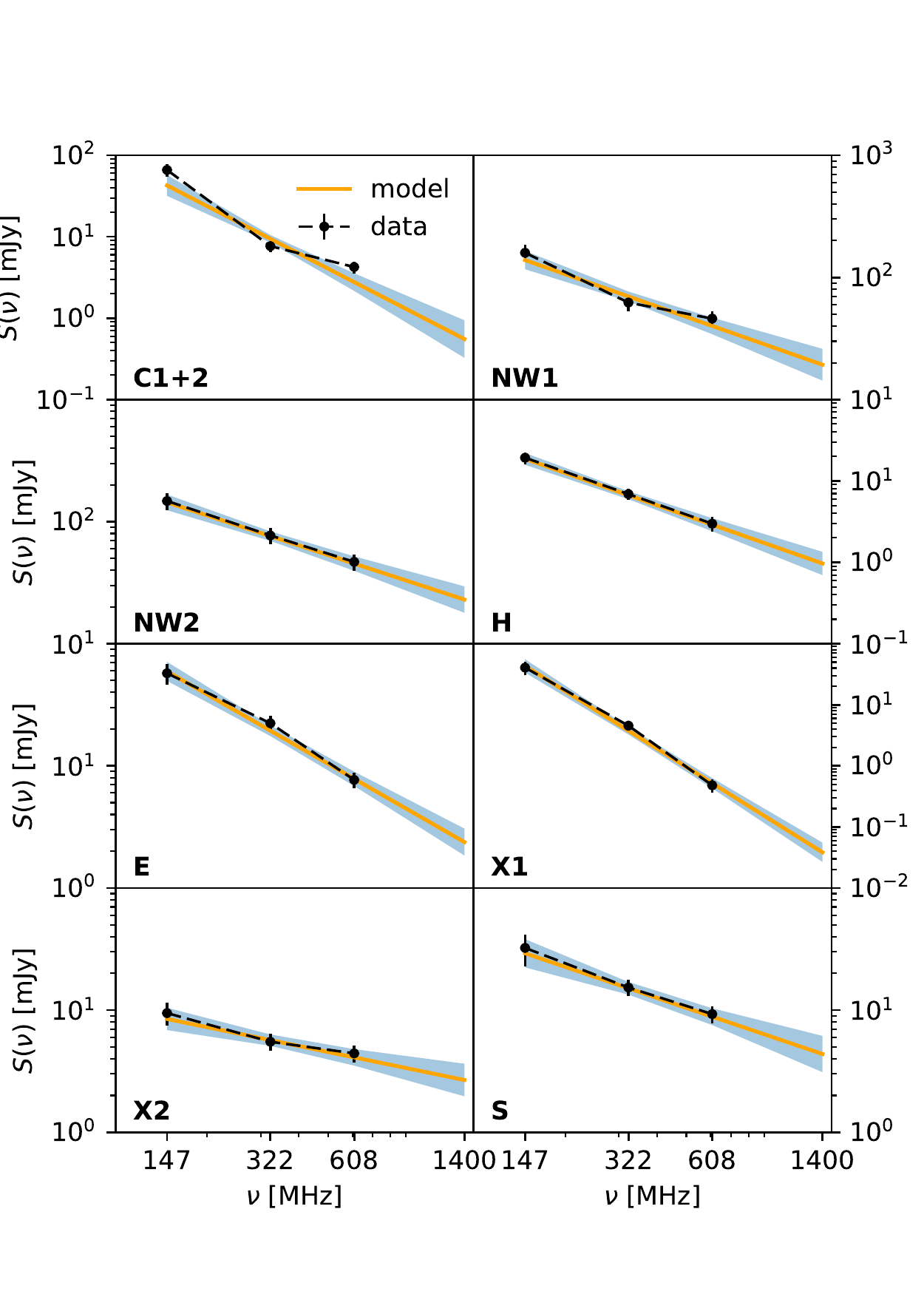}}
            \vspace{-10mm}
		\caption{The measured fluxes, $S(\nu)$, for the sources with are in blue. The mean posterior model fluxes are with orange dashed lines with $1\sigma$ shaded intervals.}
\label{fig:spxid}
\end{figure}

We modelled the flux with two parameters, the spectral index $\alpha$ and $S_{\rm 1.4~GHz}$, according to
\begin{align}
	S(\nu) = S_{\rm 1.4~GHz} \left(\frac{\nu}{\rm 1.4~GHz}\right)^\alpha.
\end{align} 
We used the Metropolis-Hastings algorithm to sample the posterior distribution of the model. 
This is a common straightforward way to sample from the posterior distribution of a model given only knowledge of the a priori distribution and a method of computing the observables.
We report the resulting posterior model parameters and uncertainties in Table~\ref{tab:spectral} with Gaussian $\alpha$ and log-normal $S_{\rm 1.4~GHz}$.
The resulting model for each source with $1\sigma$ bands is shown in Fig.~\ref{fig:spxid}. 



\begin{table*}[ht]
  \centering
	\caption{Integrated spectral fluxes, spectral index, and rest-frame $P_{\rm 1.4~GHz}$ with $\Lambda$CDM and $z=0.516$.}
\begin{tabular}{lccccccc}
    \hline
	\bf Source & $S_{150~\rm{MHz}}$ & $S_{325~\rm{MHz}}$ & $S_{610~\rm{MHz}}$ & $\alpha$ & $S_{1.4~\rm{GHz}}$ & $P_{1.4~\rm{GHz}}$ & Nature\\
	&(mJy)&(mJy)&(mJy)& & (mJy) & ($10^{24}$W/Hz) & \\
    \hline
C1+2 & $65 \pm 10$ & $8 \pm 1$ & $4.2 \pm 0.7$\tablefootmark{b} & $-1.9 \pm 0.4$ & $0.6^{+0.4}_{-0.2}$ & $0.6^{+0.3}_{-0.2}$ & lobes?\\
NW1 & $160 \pm 30$ & $62 \pm 9$ & $46 \pm 7$ & $-0.9 \pm 0.2$ & $19^{+7}_{-5}$ & $13^{+3}_{-3}$& relic?\\
NW2 & $150 \pm 20$ & $75 \pm 10$ & $47 \pm 7$ & $-0.8 \pm 0.2$ & $23^{+7}_{-5}$ & $15^{+3}_{-3}$&head-tail\\
H & $10 \pm 2$\tablefootmark{c} & $4.0 \pm 0.7$\tablefootmark{c} & $1.2 \pm 0.5$\tablefootmark{c} & $-1.2 \pm 0.4$ & $0.7^{+0.7}_{-0.3}$ & $0.5^{+0.4}_{-0.2}$&halo?\\
E & $55 \pm 10$ & $22 \pm 3$ & $8 \pm 1$ & $-1.4 \pm 0.2$ & $2.4^{+0.7}_{-0.6}$ & $2.0^{+0.4}_{-0.4}$&head-tail\\
X1 & $40 \pm 10$ & $4.6 \pm 0.7$ & $0.5 \pm 0.1$\tablefootmark{a} & $-3.1 \pm 0.3$ & $0.04^{+0.02}_{-0.01}$ & $0.07^{+0.02}_{-0.02}$&AGN?\\
X2 & $10 \pm 2$\tablefootmark{a} & $5.5 \pm 0.9$ & $4.4 \pm 0.7$ & $-0.5 \pm 0.2$ & $2.7^{+1.0}_{-0.7}$ & $1.6^{+0.4}_{-0.3}$&point?\\
S & $30 \pm 10$ & $15 \pm 2$ & $9 \pm 1$ & $-0.8 \pm 0.2$ & $4^{+2}_{-1}$ & $3.0^{+0.8}_{-0.6}$&head-tail\\
	\hline
\end{tabular}
\tablefoot{\newline
		\tablefoottext{a}{Point-source, peak flux below detection limit $5\sigma_{\rm rms}$, and aperture flux measurement.}\newline
\tablefoottext{b}{Both resolved at 610~MHz;
        $S_{610~\rm{MHz}}^{\rm C1}=2.4\pm 0.2$~mJy and $S_{610~\rm{MHz}}^{\rm C2}=1.9\pm 0.2$~mJy.}\newline
\tablefoottext{c}{Resolved source, measurement within $17~\arcsec$ circular aperture.}
    }
  \label{tab:spectral}
\end{table*}

We labelled three sources S, X1, and X2 near $R_{500}$. 
Source S has an extended radio morphology that is indicative of a tailed radio galaxy, where the tail direction implies its projected relative velocity with respect to the ICM \citepads[see for e.g.]{1974IAUS...58..109M}.
The tail of source S implies that it is moving away from the projected cluster centre.
It has a spectral index of $-0.8 \pm 0.2$, which is in line with a radio lobe.

\begin{figure}[h]
\centering
\resizebox{\columnwidth}{!}{\includegraphics{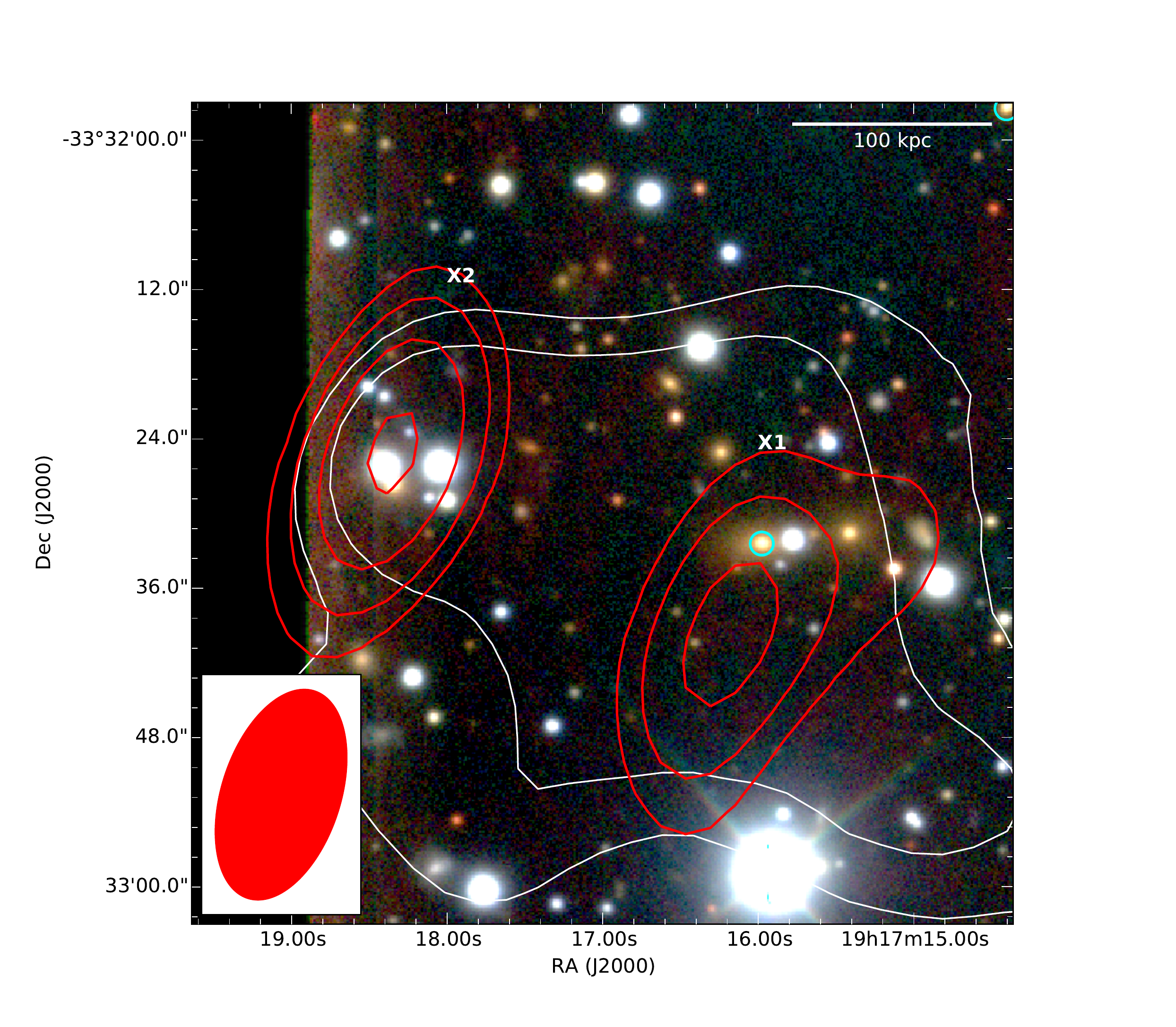}}
	\caption{X-ray (white contours) sub-nodule region with \emph{Gemini} \textit{gri}-colour image, 320~MHz contours (red), and GMOS spectroscopic sources as cyan circles.}
 \label{fig:optxnod}
\end{figure}

Sources X1 and X2 are situated above the X-ray nodule of the SE sub-cluster, and they have quite differing spectral indices of $-3.1 \pm 0.3$ and  $-0.5 \pm 0.2$, respectively. 
Figure~\ref{fig:optxnod} shows the optical region around the sources.
Red elliptical galaxies cluster strongly around source X1.
This suggests that X1 is associated with one of the giant red elliptical galaxies that make up the sub-cluster, and its steep spectral index suggests that it might be a merger-reactivated AGN.


To the east of the main cluster we have labelled an extended source, E, which has a tailed morphology similar to source S, and also a spectroscopically confirmed cluster member source near the radio surface brightness peak.
The spectral index of source E is $-1.4 \pm 0.2$, which suggests that there is some significant electron ageing.
This suggests that there might be some merger-induced reacceleration of aged electrons from a previous era of activity, or alternatively, some flux contamination from the central cluster region.
The orientation of the tail implies that source E is moving towards the projected centre of the main cluster.

Two compact sources, labelled C1 and C2, are located between source E and the cluster centre.
At 610~MHz, both patches are resolved and have similar fluxes of $2.4\pm 0.2$~mJy and $1.9\pm 0.2$~mJy, respectively. 
However, they are unresolved at 150~MHz and 325~MHz.
We therefore calculated the integrated spectral index of the sum of the two sources (in Table~\ref{tab:spectral} as C1+2), and found it to be very steep, $-1.9 \pm 0.4$.
There are spectroscopic cluster members near the two brightness peaks of C1 and C2, although it is unclear whether they are counterparts, as there is are also galaxies between C1 and C2 that are not confirmed as cluster members.
The steep spectral index supports that C1 and C2 are lobes of a galaxy.

\begin{figure}[h]
\centering
\resizebox{\columnwidth}{!}{\includegraphics{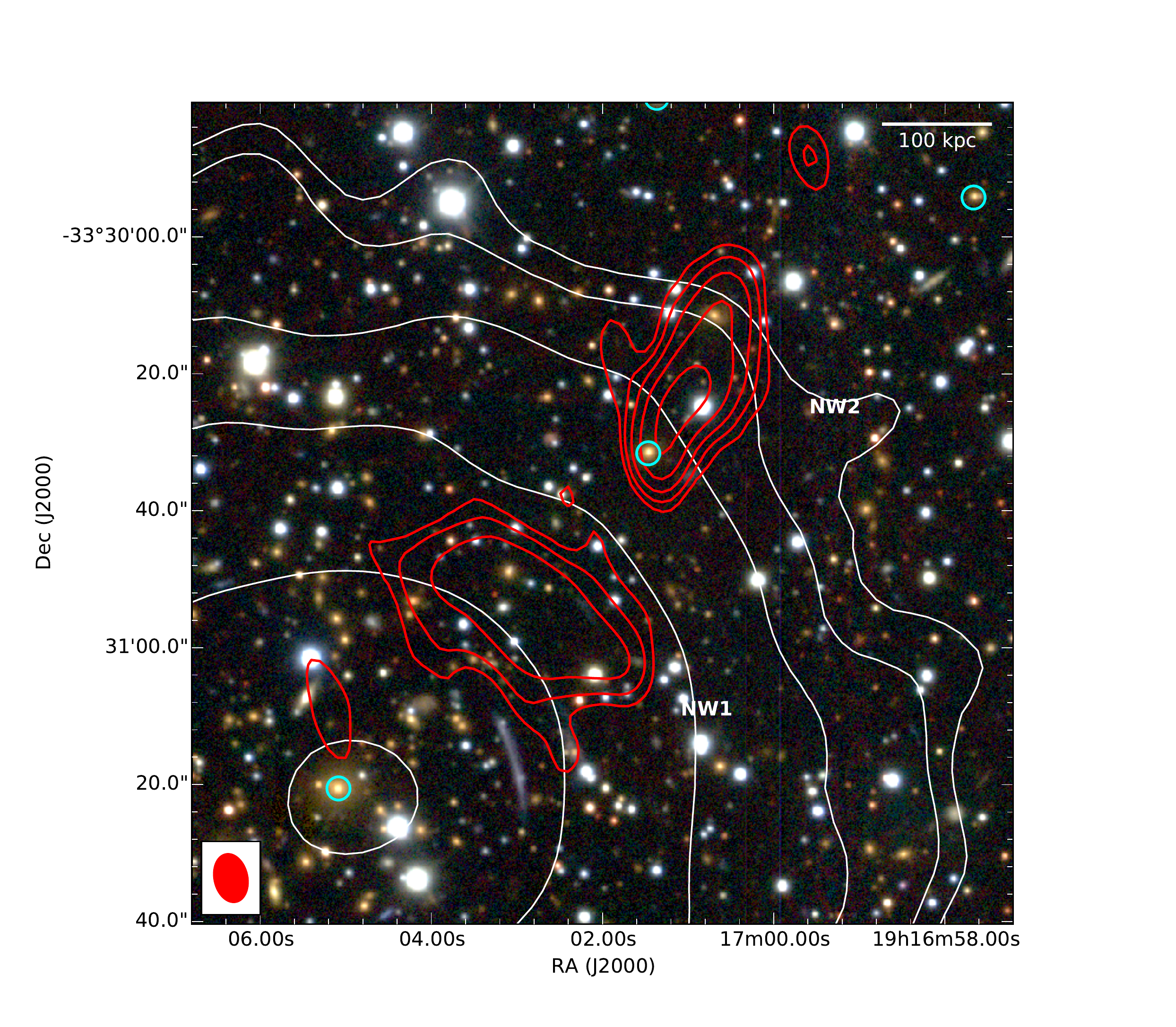}}
	\caption{\emph{Gemini} \textit{gri}-colour image with 610~MHz contours at $(5,10,20,40,80,160)\sigma_{\rm rms}$, and GMOS spectroscopic sources as cyan circles. We also show a strong lensing arc south of NW1. The brightest cluster galaxy is the southernmost cyan circle. X-ray contours are white.}
 \label{fig:radio610}
\end{figure}

To the north-west of the main cluster are two extended sources labelled NW1 and NW2.
Figure~\ref{fig:radio610} shows a zoom of this region with 610~MHz contours over the \emph{Gemini} \textit{gri}-colour image.

Source NW2 has an orientation perpendicular to the X-ray profile, with the largest linear scale (LLS) of 40\arcsec (250~kpc).
Its southernmost 610~MHz contour is 62\arcsec (380~kpc) away from the brightest cluster galaxy (BCG; $19\mathrm{h}17\mathrm{m}05.09\mathrm{s}$, $-33\mathrm{d}31\arcmin20.9\arcsec$).
There is a slight brightening at the southern tip and a nearby spectroscopically confirmed cluster member (a giant red elliptical galaxy, shown by the cyan circle in Fig.~\ref{fig:radio610}).
The integrated spectral index of NW2 is $-0.8 \pm 0.2$.

Source NW1 is oriented along the tangent of the X-ray profile, with LLS of 44\arcsec (270~kpc).
Its nearest 610~MHz contour is 28\arcsec (170~kpc) from the BCG.
The surface brightness along its length is uniform, with little variation compared to the brightness of source NW2.
The integrated spectral index is $-0.9 \pm 0.2$.

In the central region, at 150~MHz and 325~MHz, there is a diffuse bridge of radio emission between the north-western and eastern parts of the cluster, which we have labelled H because its central location and extended morphology suggest that it is a radio halo.
This source is noticeably extended in a region containing other compact sources (C1 and C2), therefore we did not use Gaussian source extraction to measure fluxes for H.
After fitting all other sources with Gaussians and subtracting them from the field we took the flux within a $17\arcsec$ circular aperture. 

We chose this aperture for two reasons.
The first is that we required the aperture to fit completely within the region between C1+2 and NW1 to avoid flux contamination.
In the 150~MHz map, the angular separation between C1+2 and NW1 is approximately 23\arcsec because of the lower resolution.
At 325~MHz the separation is approximately 45\arcsec.
The second reason is that we wish to have at least one beam of collecting area at all frequencies.
The aperture area is $900~\mathrm{arcsec}^2$, which is equivalent to 1 beam at 150~MHz, 5 beams at 325~MHz, and 23 beams at 610~MHz.
These criteria fixed the circular aperture size to 17\arcsec.
The spectral index of this extended source is measured to be $-1.2 \pm 0.4$.

\subsection{Nature of the diffuse emission}
\label{sec:dis}

In this disturbed galaxy cluster, the extended morphology and tangential orientation of source NW1 is in line with a radio relic tracing a shock front, and its spectral index is consistent with recently accelerated electrons.
The X-ray data exposure time, 14.2~ks, is too short to detect a counterpart temperature jump, and there are no polarisation data.
The projected distance of NW1 from the BCG is relatively small for a radio relic \citepads{2012MNRAS.421.3375V}, since shock structure is typically found farther out in the cluster periphery. 
A possible explanation is that we are viewing this system from a partially head-on perspective.
In addition to the creation of relics and halos, shocks in the ICM can revive fossil radio plasma from a previous episode of activity \citepads{2001A&A...366...26E} which can have filamentary structure. 
On the other hand, we might expect the spectra of NW1 to be steeper if this were the case \citepads[see e.g. sources in][]{2011A&A...527A.114V}.

Source NW2 has a morphology similar to sources E and S, with a tail direction implying it is an infalling cluster member.
The spectroscopic cluster member could be related to the slight brightening in the southern tip of NW2.
We remark that the red elliptical galaxy at the northern tip of NW2 is likely also a cluster member that might likewise contribute to the extended emission.




The diffuse source H is consistent with a radio halo, in line with the evidence that these sources are found in dynamically disturbed systems. 
{Current data suggest that the halo has a spectrum of $-1.2 \pm 0.4$, although higher sensitivity observations are required to derive better constraints. 
Measuring halo spectra at this redshift is  very important since at $z>0.5$ IC losses are expected to dominate. 
Simple modelling predicts that only one-third of very massive $\sim 1-2 \times 10^{15} M_{\odot}$ merging systems at redshift $\sim 0.5$ (such as PLCK~G004.5-19.5) are predicted to have spectra flatter than $\alpha = -1.5$, with the majority of cases generating halos with much steeper spectra \citepads{2010A&A...509A..68C}. 
}

Radio halos can also, if less frequently, have filamentary structure \citepads[e.g.][]{2005A&A...430L...5G}.
Since NW1 has a rather small projected distance on the sky from the BCG, and since we have no polarisation data and not enough data for an X-ray shock, we cannot rule out that NW1 is part of the radio halo that has survived IC losses.
However, we rate this as unlikely since we expect IC losses, which scale as $(1+z)^4$, to heavily diminish the flux at high frequencies, which leads to the USS halo nature.
The brightness of NW1 at 610~MHz indeed seems to argue against this.
Furthermore, \citetads{2014A&A...562A..43S} found very bright emission in the region of NW1 in archival 1.4~GHz NVSS.
Given that PLCK~G004.5-195 is at a mean redshift of 0.516, we would not expect remnant halo flux to be bright at higher frequencies.






\section{Conclusion}


We have performed high-resolution 610~MHz, 325~MHz, and 150~MHz observations of the cluster PLCK~G004.5-19.5 with the GMRT.
We found a complex arrangement of central and peripheral diffuse emission, as well as several head-tail galaxies.

We reprocessed archival \emph{XMM-Newton} data and found morphological parameters that imply that PLCK~G004.5-19.5 is a disturbed system \citepads{2010ApJ...721L..82C}.
Furthermore, a previously unidentified south-eastern sub-cluster is observed with X-ray unabsorbed luminosity $(6.7 \pm 1.4) \times 10^{43}~\mathrm{erg}\,s^{-1}$, approximately a factor 14--16 lower than the luminosity of them main cluster in the same aperture. 

{Most remarkably, we found a partially face-on radio relic (NW1) and centrally located radio halo (H) with spectral index ($-1.2 \pm 0.4$).
Our interpretation of this finding favours the dynamically young state suggested by the X-ray morphology parameters, as well as the morphology of the radio emission.}

PLCK~G004.5-19.5 is the fifth cluster beyond $z=0.5$ known to host a radio relic, and it is one of the few known instances of a ($z>0.5$) {radio halo}.

\begin{acknowledgements}
J.\ A., A.\ S., and H.\ R. acknowledge financial support from an NWO top subsidy (614.001.006). 
C.\ S. acknowledges support from the European Research Council under FP7 grant number 279396.
We thank the staff of the GMRT, who have made these observations possible. 
	The GMRT is run by the National Centre for Radio Astrophysics of the Tata Institute of Fundamental Research.
	This work is based on observations obtained with XMM-Newton, an ESA science mission with instruments and contributions directly funded by ESA member states and the USA (NASA). 
	The SRON Netherlands Institute for Space Research is supported financially by NWO, the Netherlands Organisation for Space Research.
	G.\ B. acknowledges partial support from grant PRIN-INAF2014.
	J.\ A. thanks R.\ Cassano for private discussion on use of morphological parameters at higher redshifts.
\end{acknowledgements}

\bibliographystyle{aa}
\bibliography{cite,Plck_part_Francois}

\begin{thebibliography}{41}
\expandafter\ifx\csname natexlab\endcsname\relax\def\natexlab#1{#1}\fi

\bibitem[{{Bonafede} {et~al.}(2012){Bonafede}, {Br{\"u}ggen}, {van Weeren},
  {Vazza}, {Giovannini}, {Ebeling}, {Edge}, {Hoeft}, \&
  {Klein}}]{2012MNRAS.426...40B}
{Bonafede}, A., {Br{\"u}ggen}, M., {van Weeren}, R., {et~al.} 2012, \mnras,
  426, 40

\bibitem[{{Bonafede} {et~al.}(2009){Bonafede}, {Feretti}, {Giovannini},
  {Govoni}, {Murgia}, {Taylor}, {Ebeling}, {Allen}, {Gentile}, \&
  {Pihlstr{\"o}m}}]{2009A&A...503..707B}
{Bonafede}, A., {Feretti}, L., {Giovannini}, G., {et~al.} 2009, \aap, 503, 707

\bibitem[{{Botteon} {et~al.}(2016){Botteon}, {Gastaldello}, {Brunetti}, \&
  {Kale}}]{2016MNRAS.tmp.1213B}
{Botteon}, A., {Gastaldello}, F., {Brunetti}, G., \& {Kale}, R. 2016, \mnras
  [\eprint[arXiv]{1607.04641}]

\bibitem[{Briggs(1995)}]{briggs}
Briggs, D. 1995, PhD thesis, New Mexico Institute of Mining Technology

\bibitem[{{Brunetti}(2016)}]{2016PPCF...58a4011B}
{Brunetti}, G. 2016, Plasma Physics and Controlled Fusion, 58, 014011

\bibitem[{{Brunetti} {et~al.}(2008){Brunetti}, {Giacintucci}, {Cassano},
  {Lane}, {Dallacasa}, {Venturi}, {Kassim}, {Setti}, {Cotton}, \&
  {Markevitch}}]{2008Natur.455..944B}
{Brunetti}, G., {Giacintucci}, S., {Cassano}, R., {et~al.} 2008, \nat, 455, 944

\bibitem[{{Brunetti} \& {Jones}(2014)}]{2014IJMPD..2330007B}
{Brunetti}, G. \& {Jones}, T.~W. 2014, International Journal of Modern Physics
  D, 23, 1430007

\bibitem[{{Cassano} {et~al.}(2010{\natexlab{a}}){Cassano}, {Brunetti},
  {R{\"o}ttgering}, \& {Br{\"u}ggen}}]{2010A&A...509A..68C}
{Cassano}, R., {Brunetti}, G., {R{\"o}ttgering}, H.~J.~A., \& {Br{\"u}ggen}, M.
  2010{\natexlab{a}}, \aap, 509, A68

\bibitem[{{Cassano} {et~al.}(2006){Cassano}, {Brunetti}, \&
  {Setti}}]{2006MNRAS.369.1577C}
{Cassano}, R., {Brunetti}, G., \& {Setti}, G. 2006, \mnras, 369, 1577

\bibitem[{{Cassano} {et~al.}(2010{\natexlab{b}}){Cassano}, {Ettori},
  {Giacintucci}, {Brunetti}, {Markevitch}, {Venturi}, \&
  {Gitti}}]{2010ApJ...721L..82C}
{Cassano}, R., {Ettori}, S., {Giacintucci}, S., {et~al.} 2010{\natexlab{b}},
  \apjl, 721, L82

\bibitem[{{Cornwell} {et~al.}(2008){Cornwell}, {Golap}, \&
  {Bhatnagar}}]{2008ISTSP...2..647C}
{Cornwell}, T.~J., {Golap}, K., \& {Bhatnagar}, S. 2008, IEEE Journal of
  Selected Topics in Signal Processing, 2, 647

\bibitem[{{De Luca} \& {Molendi}(2004)}]{2004A&A...419..837D}
{De Luca}, A. \& {Molendi}, S. 2004, \aap, 419, 837

\bibitem[{{Donnert} {et~al.}(2013){Donnert}, {Dolag}, {Brunetti}, \&
  {Cassano}}]{2013MNRAS.429.3564D}
{Donnert}, J., {Dolag}, K., {Brunetti}, G., \& {Cassano}, R. 2013, \mnras, 429,
  3564

\bibitem[{{Drury}(1983)}]{1983RPPh...46..973D}
{Drury}, L.~O. 1983, Reports on Progress in Physics, 46, 973

\bibitem[{{En{\ss}lin} \& {Gopal-Krishna}(2001)}]{2001A&A...366...26E}
{En{\ss}lin}, T.~A. \& {Gopal-Krishna}. 2001, \aap, 366, 26

\bibitem[{{Freudling} {et~al.}(2013){Freudling}, {Romaniello}, {Bramich},
  {Ballester}, {Forchi}, {Garc{\'{\i}}a-Dabl{\'o}}, {Moehler}, \&
  {Neeser}}]{2013A&A...559A..96F}
{Freudling}, W., {Romaniello}, M., {Bramich}, D.~M., {et~al.} 2013, \aap, 559,
  A96

\bibitem[{{Govoni} {et~al.}(2005){Govoni}, {Murgia}, {Feretti}, {Giovannini},
  {Dallacasa}, \& {Taylor}}]{2005A&A...430L...5G}
{Govoni}, F., {Murgia}, M., {Feretti}, L., {et~al.} 2005, \aap, 430, L5

\bibitem[{{Healey} {et~al.}(2007){Healey}, {Romani}, {Taylor}, {Sadler},
  {Ricci}, {Murphy}, {Ulvestad}, \& {Winn}}]{2007ApJS..171...61H}
{Healey}, S.~E., {Romani}, R.~W., {Taylor}, G.~B., {et~al.} 2007, \apjs, 171,
  61

\bibitem[{{Intema}(2014)}]{2014ascl.soft08006I}
{Intema}, H.~T. 2014, {SPAM: Source Peeling and Atmospheric Modeling},
  Astrophysics Source Code Library

\bibitem[{{Intema} {et~al.}(2016){Intema}, {Jagannathan}, {Mooley}, \&
  {Frail}}]{2016arXiv160304368I}
{Intema}, H.~T., {Jagannathan}, P., {Mooley}, K.~P., \& {Frail}, D.~A. 2016,
  ArXiv e-prints [\eprint[arXiv]{1603.04368}]

\bibitem[{{Intema} {et~al.}(2009){Intema}, {van der Tol}, {Cotton}, {Cohen},
  {van Bemmel}, \& {R{\"o}ttgering}}]{2009A&A...501.1185I}
{Intema}, H.~T., {van der Tol}, S., {Cotton}, W.~D., {et~al.} 2009, \aap, 501,
  1185

\bibitem[{{Kang} \& {Ryu}(2013)}]{2013ApJ...764...95K}
{Kang}, H. \& {Ryu}, D. 2013, \apj, 764, 95

\bibitem[{{Kurtz} \& {Mink}(1998)}]{1998PASP..110..934K}
{Kurtz}, M.~J. \& {Mink}, D.~J. 1998, \pasp, 110, 934

\bibitem[{{Lindner} {et~al.}(2014){Lindner}, {Baker}, {Hughes}, {Battaglia},
  {Gupta}, {Knowles}, {Marriage}, {Menanteau}, {Moodley}, {Reese}, \&
  {Srianand}}]{2014ApJ...786...49L}
{Lindner}, R.~R., {Baker}, A.~J., {Hughes}, J.~P., {et~al.} 2014, \apj, 786, 49

\bibitem[{{Malkov} \& {O'C Drury}(2001)}]{2001RPPh...64..429M}
{Malkov}, M.~A. \& {O'C Drury}, L. 2001, Reports on Progress in Physics, 64,
  429

\bibitem[{{Maughan} {et~al.}(2008){Maughan}, {Jones}, {Forman}, \& {Van
  Speybroeck}}]{2008ApJS..174..117M}
{Maughan}, B.~J., {Jones}, C., {Forman}, W., \& {Van Speybroeck}, L. 2008,
  \apjs, 174, 117

\bibitem[{{Mernier} {et~al.}(2015){Mernier}, {de Plaa}, {Lovisari}, {Pinto},
  {Zhang}, {Kaastra}, {Werner}, \& {Simionescu}}]{2015A&A...575A..37M}
{Mernier}, F., {de Plaa}, J., {Lovisari}, L., {et~al.} 2015, \aap, 575, A37

\bibitem[{{Miley} {et~al.}(1974){Miley}, {van der Laan}, \&
  {Wellington}}]{1974IAUS...58..109M}
{Miley}, G.~K., {van der Laan}, H., \& {Wellington}, K.~J. 1974, in IAU
  Symposium, Vol.~58, The Formation and Dynamics of Galaxies, ed. J.~R.
  {Shakeshaft}, 109

\bibitem[{{Nuza} {et~al.}(2012){Nuza}, {Hoeft}, {van Weeren}, {Gottl{\"o}ber},
  \& {Yepes}}]{2012MNRAS.420.2006N}
{Nuza}, S.~E., {Hoeft}, M., {van Weeren}, R.~J., {Gottl{\"o}ber}, S., \&
  {Yepes}, G. 2012, \mnras, 420, 2006

\bibitem[{{Planck Collaboration} {et~al.}(2014){Planck Collaboration}, {Ade},
  {Aghanim}, {Armitage-Caplan}, {Arnaud}, {Ashdown}, {Atrio-Barandela},
  {Aumont}, {Aussel}, {Baccigalupi}, \& et~al.}]{2014A&A...571A..29P}
{Planck Collaboration}, {Ade}, P.~A.~R., {Aghanim}, N., {et~al.} 2014, \aap,
  571, A29

\bibitem[{{Planck Collaboration} {et~al.}(2011){Planck Collaboration},
  {Aghanim}, {Arnaud}, {Ashdown}, {Aumont}, {Baccigalupi}, {Balbi}, {Banday},
  {Barreiro}, {Bartelmann}, \& et~al.}]{2011A&A...536A...9P}
{Planck Collaboration}, {Aghanim}, N., {Arnaud}, M., {et~al.} 2011, \aap, 536,
  A9

\bibitem[{{Poole} {et~al.}(2006){Poole}, {Fardal}, {Babul}, {McCarthy},
  {Quinn}, \& {Wadsley}}]{2006MNRAS.373..881P}
{Poole}, G.~B., {Fardal}, M.~A., {Babul}, A., {et~al.} 2006, \mnras, 373, 881

\bibitem[{{Reiprich} \& {B{\"o}hringer}(2002)}]{2002ApJ...567..716R}
{Reiprich}, T.~H. \& {B{\"o}hringer}, H. 2002, \apj, 567, 716

\bibitem[{{Riseley} {et~al.}(2017){Riseley}, {Scaife}, {Wise}, \&
  {Clarke}}]{2017A&A...597A..96R}
{Riseley}, C.~J., {Scaife}, A.~M.~M., {Wise}, M.~W., \& {Clarke}, A.~O. 2017,
  \aap, 597, A96

\bibitem[{{Roettiger} {et~al.}(1999){Roettiger}, {Stone}, \&
  {Burns}}]{1999ApJ...518..594R}
{Roettiger}, K., {Stone}, J.~M., \& {Burns}, J.~O. 1999, \apj, 518, 594

\bibitem[{{Santos} {et~al.}(2008){Santos}, {Rosati}, {Tozzi}, {B{\"o}hringer},
  {Ettori}, \& {Bignamini}}]{2008A&A...483...35S}
{Santos}, J.~S., {Rosati}, P., {Tozzi}, P., {et~al.} 2008, \aap, 483, 35

\bibitem[{{Sif{\'o}n} {et~al.}(2014){Sif{\'o}n}, {Menanteau}, {Hughes},
  {Carrasco}, \& {Barrientos}}]{2014A&A...562A..43S}
{Sif{\'o}n}, C., {Menanteau}, F., {Hughes}, J.~P., {Carrasco}, M., \&
  {Barrientos}, L.~F. 2014, \aap, 562, A43

\bibitem[{{van Haarlem} {et~al.}(2013){van Haarlem}, {Wise}, {Gunst}, {Heald},
  {McKean}, {Hessels}, {de Bruyn}, {Nijboer}, {Swinbank}, {Fallows},
  {Brentjens}, {Nelles}, {Beck}, {Falcke}, {Fender}, {H{\"o}randel},
  {Koopmans}, {Mann}, {Miley}, {R{\"o}ttgering}, {Stappers}, {Wijers},
  {Zaroubi}, {van den Akker}, {Alexov}, {Anderson}, {Anderson}, {van Ardenne},
  {Arts}, {Asgekar}, {Avruch}, {Batejat}, {B{\"a}hren}, {Bell}, {Bell}, {van
  Bemmel}, {Bennema}, {Bentum}, {Bernardi}, {Best}, {B{\^i}rzan}, {Bonafede},
  {Boonstra}, {Braun}, {Bregman}, {Breitling}, {van de Brink}, {Broderick},
  {Broekema}, {Brouw}, {Br{\"u}ggen}, {Butcher}, {van Cappellen}, {Ciardi},
  {Coenen}, {Conway}, {Coolen}, {Corstanje}, {Damstra}, {Davies}, {Deller},
  {Dettmar}, {van Diepen}, {Dijkstra}, {Donker}, {Doorduin}, {Dromer}, {Drost},
  {van Duin}, {Eisl{\"o}ffel}, {van Enst}, {Ferrari}, {Frieswijk}, {Gankema},
  {Garrett}, {de Gasperin}, {Gerbers}, {de Geus}, {Grie{\ss}meier}, {Grit},
  {Gruppen}, {Hamaker}, {Hassall}, {Hoeft}, {Holties}, {Horneffer}, {van der
  Horst}, {van Houwelingen}, {Huijgen}, {Iacobelli}, {Intema}, {Jackson},
  {Jelic}, {de Jong}, {Juette}, {Kant}, {Karastergiou}, {Koers}, {Kollen},
  {Kondratiev}, {Kooistra}, {Koopman}, {Koster}, {Kuniyoshi}, {Kramer},
  {Kuper}, {Lambropoulos}, {Law}, {van Leeuwen}, {Lemaitre}, {Loose}, {Maat},
  {Macario}, {Markoff}, {Masters}, {McFadden}, {McKay-Bukowski}, {Meijering},
  {Meulman}, {Mevius}, {Middelberg}, {Millenaar}, {Miller-Jones}, {Mohan},
  {Mol}, {Morawietz}, {Morganti}, {Mulcahy}, {Mulder}, {Munk}, {Nieuwenhuis},
  {van Nieuwpoort}, {Noordam}, {Norden}, {Noutsos}, {Offringa}, {Olofsson},
  {Omar}, {Orr{\'u}}, {Overeem}, {Paas}, {Pandey-Pommier}, {Pandey}, {Pizzo},
  {Polatidis}, {Rafferty}, {Rawlings}, {Reich}, {de Reijer}, {Reitsma},
  {Renting}, {Riemers}, {Rol}, {Romein}, {Roosjen}, {Ruiter}, {Scaife}, {van
  der Schaaf}, {Scheers}, {Schellart}, {Schoenmakers}, {Schoonderbeek},
  {Serylak}, {Shulevski}, {Sluman}, {Smirnov}, {Sobey}, {Spreeuw}, {Steinmetz},
  {Sterks}, {Stiepel}, {Stuurwold}, {Tagger}, {Tang}, {Tasse}, {Thomas},
  {Thoudam}, {Toribio}, {van der Tol}, {Usov}, {van Veelen}, {van der Veen},
  {ter Veen}, {Verbiest}, {Vermeulen}, {Vermaas}, {Vocks}, {Vogt}, {de Vos},
  {van der Wal}, {van Weeren}, {Weggemans}, {Weltevrede}, {White}, {Wijnholds},
  {Wilhelmsson}, {Wucknitz}, {Yatawatta}, {Zarka}, {Zensus}, \& {van
  Zwieten}}]{2013A&A...556A...2V}
{van Haarlem}, M.~P., {Wise}, M.~W., {Gunst}, A.~W., {et~al.} 2013, \aap, 556,
  A2

\bibitem[{{van Weeren} {et~al.}(2011){van Weeren}, {R{\"o}ttgering}, \&
  {Br{\"u}ggen}}]{2011A&A...527A.114V}
{van Weeren}, R.~J., {R{\"o}ttgering}, H.~J.~A., \& {Br{\"u}ggen}, M. 2011,
  \aap, 527, A114

\bibitem[{{van Weeren} {et~al.}(2009){van Weeren}, {R{\"o}ttgering},
  {Br{\"u}ggen}, \& {Cohen}}]{2009A&A...505..991V}
{van Weeren}, R.~J., {R{\"o}ttgering}, H.~J.~A., {Br{\"u}ggen}, M., \& {Cohen},
  A. 2009, \aap, 505, 991

\bibitem[{{Vazza} {et~al.}(2012){Vazza}, {Br{\"u}ggen}, {Gheller}, \&
  {Brunetti}}]{2012MNRAS.421.3375V}
{Vazza}, F., {Br{\"u}ggen}, M., {Gheller}, C., \& {Brunetti}, G. 2012, \mnras,
  421, 3375

\end{thebibliography}

\end{document}